# Horizontal Attacks against ECC: from Simulations to ASIC

Ievgen Kabin[1], Zoya Dyka[1], Dan Klann[1] and Peter Langendoerfer[1]

[1] *IHP – Leibniz-Institut für innovative Mikroelektronik*, Frankfurt (Oder), Germany
`{kabin, dyka, klann, langendoerfer}@ihp-microelectronics.com`

**Abstract.** In this paper we analyse the impact of different compile options on the success rate of side channel analysis attacks. We run horizontal differential side channel attacks against simulated power traces for the same *kP* design synthesized using two different compile options after synthesis and after layout. As we are interested in the effect on the produced ASIC we also run the same attack against measured power traces after manufacturing the ASIC. We found that the *compile_ultra* option reduces the success rate significantly from 5 key candidates with a correctness of between 75 and 90 per cent down to 3 key candidates with a maximum success rate of 72 per cent compared to the simple *compile* option. Also the success rate after layout shows a very high correlation with the one obtained attacking the measured power and electromagnetic traces, i.e. the simulations are a good indicator of the resistance of the ASIC.

## 1 Introduction

The Internet of Things (IoT) is demanding of proper security features such confidentiality and integrity even for resource constraint devices such as embedded devices or wireless sensor nodes. Due to the scarce resources software implementations especially for asymmetric cryptographic approaches are not applicable, so hardware accelerators are becoming a must. It is also part of the very nature of IoT that some devices are deployed in the wild at difficult to access place and are connected wireless. This means potential attackers can gain physical access to those devices. If this happens they can measure parameters such as power consumption or electromagnetic emanation. The analysis of these measurements may reveal the used private key and by that render the deployed security solution useless.

In order to cope with the above mentioned issue implementations of cryptographic operations need to be side channel analysis attack (SCA) resistant. This is a very challenging task as the large amount of publications reporting on successful attacks impressively proofs. The best if not the only way to get an SCA resistant implementation is to run SCA attacks against your implementation at the earliest possible stage. This is kind of common sense and how to do it was discussed in several publications already. But to the best of our knowledge this paper is the first one presenting a thorough analysis of the influence of design tool in the steps from a behavioural model down to an ASIC. The major contributions of this paper are:

— Discussion of the success rate of horizontal attacks against an ECC design after synthesis and layout for different design compiler options.
— Discussion and comparison of the success rate on simulated power traces with those measured on the ASIC manufactured using the more promising compile option.
— Discussion and comparison of the success rate on measured electromagnetic traces with the ones on simulated and measured power traces. The correlation between the clock cycles in which the success rate is high is pretty good which indicates that at least in this case the simulation after layout provide also some indication about potential issues when attacked using EM traces. This is of high importance as detailed EM simulations for that large designs are currently infeasible.
— Our results show that some compile option are helping to reduce side channel leakage while other don't. But be aware that compile options alone won't make a bad design SCA resistant.

The rest of this paper is structured as follows. In section 2 we give a brief overview of different attacks classifications. In section 3 we explain the implementation details of the investigated design. In section 4 we present in detail how we performed the horizontal DPA attack on simulated power traces using *the comparison to the mean* [1] method. In section 5 we give an information about produced ASIC. In section 6 we discuss the results of the performed attacks using measured and simulated traces. The paper finishes with short conclusions.

## 2    Background: verical and horizontal attacks

One of most often used known classification of SCA attacks is the classification into simple and differential attacks, for example simple power analysis (SPA) and differential power analysis (DPA). If the secret (private) key can be successful revealed using a single power trace without any use of statistical methods the attack is classified as SPA attack. For DPA attacks [1] a lot of traces with different inputs are required for revealing the secret key using statistical analysis of the traces. If for revealing the key correlation coefficients are calculated the attack is usually classified as a correlation power analysis (CPA) attack [2]. Attacks known as collision attacks [3]-[5] need usually more than one trace for the analysis but not so many as the classical DPA attacks.

In 2010 Clavier et al. [6] applied the correlation analysis to reveal the secret exponent of an RSA implementation using a single power trace. They called this attack horizontal correlation analysis (HCA) attack. Due to the diversity of the attack classifications they introduced the simplified classification of SCA attacks into horizontal and vertical attacks. Corresponding to their classification horizontal attacks are single-trace attacks, for example SPA or HCA attacks. Practical examples of horizontal attacks against asymmetric cryptographic approaches such RSA or ECC are the SPA described in [7], simple electromagnetic analysis (SEMA) attacks e.g. [8], the Big Mac attack [9], the localized EMA attack [10], horizontal collision correlation analysis (HCCA) attacks [11]-[12] , horizontal DPA and DEMA attacks [13], [14]. Vertical attacks corresponding to the new classification are "more than one trace" attacks i.e. classical DPA, CPA and collision-based attacks.

---

[1] we called this method in our earlier papers "the difference of the mean" as it differs from the well-known statistical method "*difference of means*". To avoid confusions and/or misunderstandings due to the very similar names we call it from now on *the comparison to the mean*.

Vertical attacks exploit the fact that there is some kind of relation between the different inputs that are known by the attacker and the secret but constant key. The well-known countermeasures such as elliptic curve (EC) point blinding, randomization of projective coordinates of EC points or the key randomization proposed in [15] are effective against most vertical attacks because they resolve the knowledge of the attacker about the processed inputs or make the processed secret value no longer constant. But these countermeasures are not effective against horizontal attacks [6], [16] and against the vertical address bit DPA attack introduced in 2003 in [17]. This address bit DPA exploited key dependent addressing of registers in the Montgomery $kP$ algorithm. 1000 traces were analysed to reveal the key. The horizontal address bit DPA reported in [13] successfully revealed the secret scalar analyzing only a single trace of a $kP$ execution. Addressing of registers and other ECC design blocks caused the leakage exploited in the attack.

Most industrial authentication products such as NXP A1006 [18] and Infineon SLE95250 [19] are based on the Montgomery $kP$ algorithm. So, countermeasures against the vertical and horizontal address bit DPA such as the randomized addressing of registers [17] or the randomization of the main loop of the Montgomery $kP$ algorithm [20] or developing a regular schedule in which the blocks are addressed [21] are of high commercial interest and have to be implemented. The activity of the field multiplier can increase the inherent resistance of $kP$ designs against SCA attacks. To exploit the field multiplier as a countermeasure different multiplication methods have to be combined [22]. We didn't find any information about the implementation of such a countermeasure in industrial chips. We found only classical countermeasures such as the randomization of coordinates of EC points were reported [23].

## 3 Investigated ECC Designs

### 3.1 Structure of the implemented $kP$ designs

The implemented ECC design is a hardware accelerator for the multiplication of an elliptic curve point $P$ with a scalar $k$. The operation is denoted as $kP$. The structure of the implemented $kP$-hardware accelerator is the same for designs investigated in this paper and it is shown on **Fig. 1**.

The $kP$ accelerator obtains a scalar $k$ and two affine coordinates $x$ and $y$ of a point $P$ of the EC B-233 [24] as inputs for processing. The numbers $x$, $y$ and $k$ are up to 233 bit long binary numbers that represent elements of $GF(2^{233})$ with the irreducible polynomial $f(t)=t^{233}+t^{74}+1$.

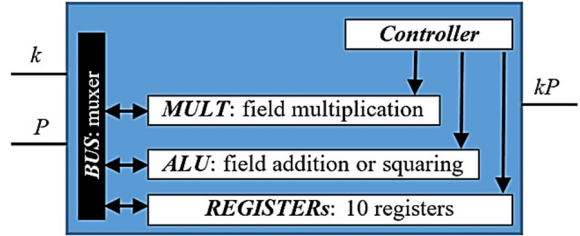

**Fig. 1.** Structure of the implemented $kP$-accelerator.

The hardware accelerator processes the scalar $k$ bitwise according to the Montgomery $kP$ algorithm using Lopez-Dahab projective coordinates [25]. Our implementations are based on Algorithm 2 presented in [26]. This implementation allows to perform all operations in parallel to the field multiplication.

### 3.2 Field multiplier

The multiplication is the most complex field operation in our designs. The polynomial multiplication (i.e. the first step of the multiplication of elements of in $GF(2^n)$) can be realized by applying a lot of multiplication methods e.g. the classical, Karatsuba [27], Winograd [28] or combinations of these [29].

We implemented the field multiplier using the 4-segment Karatsuba multiplication method according to a fixed calculation plan as described in [30]. The structure of our field multiplier is shown in **Fig. 2**.

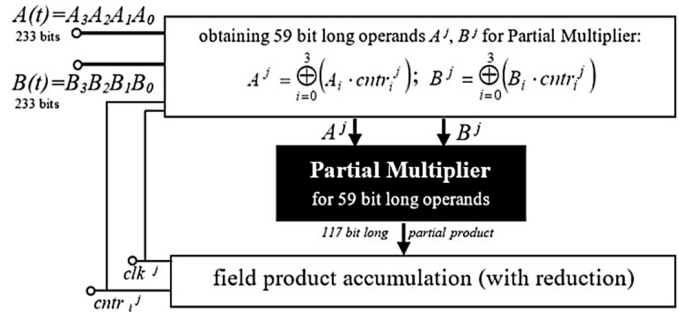

**Fig. 2.** Structure of the field multiplier.

Two up to 233 bit long operands $A(t)$ and $B(t)$ are segmented into four parts: $A_3$, $A_2$, $A_1$, $A_0$ and $B_3$, $B_2$, $B_1$, $B_0$ respectively. The parts $A_3$ and $B_3$ are 56 bit long. All other parts are 59 bit long. The field multiplier takes 9 clock cycles ($clk^j$, $j=0,1,2,...,8$) to calculate the product of 233 bit long operands. In each clock cycle only one of 9 partial products of the 59 bit long operands is calculated accordingly to the 4 segment iterative Karatsuba multiplication method. The signals from Controller $cntr^i_j$ organize the calculation of operands for the Partial Multiplier clockwise. All partial products are accumulated in a register of the multiplier. The field product will be accumulated iteratively, step-by-step (or clock-by-clock), using the calculated partial products. The reduction could be performed only once, after all 9 partial multiplications, but the reduction consumes energy, i.e. it is a kind of "dummy operation" that hides partially the activity of other blocks of the $kP$ design. Thus, performing of the reduction clockwise can increase the resistance of the $kP$ design against SCA attacks. Even though we did not yet evaluate this effect, we perform the reduction in each clock cycle.

A partial multiplier was implemented using the classical multiplication formula only, i.e. it implements the following formula:

$$C = A \cdot B = \sum_{i=0}^{2n-2} c_i \cdot t^i, \text{ with } c_i = \bigoplus_{i=k+l} a_k \cdot b_l, \forall k, l < n \qquad (1)$$

Here $n$=59 is the length of the partial multiplicands.

The gate complexity of this multiplier, i.e. the amount of AND and XOR gates which are necessary to implement its functionality corresponding to formula (1) is $n^2$ of AND gates and $(n-1)^2$ of XOR gates, i.e.: $GC_{n=59}=3481_\& + 3364_{XOR}$.

We synthesized the $kP$ design using the gate library for our 250 nm technology for a clock cycle of 50 ns. We used two compilation techniques in the Synopsys Design Compiler (version K-2015.06-SP2) [31] to perform hardware optimizations: simple *compile* option and *compile_ultra* without clock gating, with *–no-autoungroup* option to preserve all hierarchies in the design. In the rest of paper the design obtained with a simple *compile* option is called as *D_noUltra*. The design obtained by using the *compile_ultra* option is called *D_ultra*.

The *compile* command performs logic-level and gate-level synthesis and optimization of the current design. The optimization process trades off timing and area constraints to provide the smallest possible circuit that meets the specified timing requirements. Values for area and speed of components used in synthesizing and optimizing the design are obtained from user-specified libraries.

The *compile_ultra* command performs a high-effort compilation on the design to achieve an optimum quality of results by enabling all DC Ultra features, therefore increasing the performance and significantly reducing design area and power consumption. It is targeted toward high-performance designs with very tight timing constraints.

Both designs investigated here, i.e. *D_noUltra* and *D_ultra*, require the same amount of clock cycles (about 13000) for a single $kP$ operation. We simulated the power consumption of the $kP$ designs obtained after syntheses and layout while the $kP$ operation was executed using the same input data. All $kP$ traces were simulated using the Synopsis PrimeTime suite [31].

**Table 1** presents the parameters of the investigated $kP$ designs. The goal is to show that the main parameters of the investigated $kP$ designs after syntheses and after layout such as area and power are quite different.

**Table 1.** Parameters of investigated *kP* designs

| | | Investigated *kP* designs | | | Field multiplier | | | |
| | | | | | *area* | | *power* | |
| | *name* | *area, mm²* | *power, mW* | *total, mm²* | *relative to kP design* | *total, mW* | *relative to kP design* |
|---|---|---|---|---|---|---|---|
| *Synthesis* | *D_noUltra* | 1.78 | 21.1 | 0.69 | 37.4 % | 14.6 | 69.2 % |
| | *D_ultra* | 1.68 | 13.7 | 0.62 | 36.8 % | 8.58 | 62.5 % |
| *Layout* | *D_noUltra* | 1.77 | 25.5 | 0.67 | 37.6 % | 16 | 62.7 % |
| | *D_ultra* | 1.70 | 16.6 | 0.63 | 37.0 % | 8.5 | 51.1 % |

## 4    Performed attack using the comparison to the mean

The simulated traces are noiseless. Due to this fact we compressed the simulated power trace, i.e. we represented each clock cycle using only one value – the average power value of the clock cycle. **Fig. 3**(*a*) and **Fig. 3**(*b*) show compressed simulated traces obtained after syntheses and layout respectively. Power traces for *D_noUltra* are shown in red and power traces for *D_ultra* in blue. It can be seen that the design obtained using the *compile_ultra* technique consumes less power.

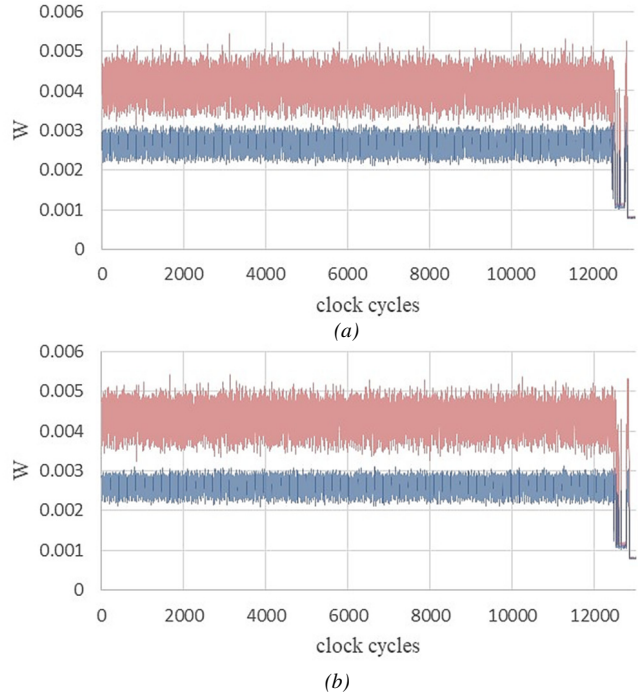

**Fig. 3.** Compressed simulated power traces for both designs: *(a)* – the traces were simulated after synthesis; *(b)* - the traces were simulated after layout.

We applied *the comparison to the mean* for statistical analysis of the traces. To perform a horizontal DPA attack we fragmented the compressed trace into slots. The compressed trace consists of *l-2=230* slots with similar profiles, where *l* is the length of the processed scalar *k*. Each slot corresponds to the processing of a key bit $k_i$ with *0≤i≤l-3* in the main loop of the implemented algorithm and consists of 54 values (one value per clock cycle). Thus, each value of the compressed trace can be represented as $v_i{}^j$, where *i* is the number of the slot and *j* is the number of the clock cycle within the slot, with *0≤i≤l-3, 1≤j≤54*. We calculated the *mean* slot, i.e. the arithmetical mean of all values with the same number *j* and different number *i*: $\overline{v^j} = \dfrac{1}{l-2}\sum_{i=0}^{l-3} v_i^j$ . Thus, the 54 calculated values $\overline{v^j}$ define the *mean* slot. We obtained 54 key candidates – one per clock cycle *j* – using the following assumption: $k_{candidate}{}^j{}_i = 1$ if $\overline{v^j} \geq v_i^j$ else $k_{candidate}{}^j{}_i = 0$ . To evaluate the success of the attack we compared the extracted key candidates with the scalar *k* that was really processed. For each key candidate we calculated its relative correctness as follows:

$$\delta_i = \frac{\#correct\_extracted\_bits(k_{candidate}{}^j{}_i)}{l-2} \cdot 100\% \qquad (2)$$

A correctness close to 0 percent means that our assumption is wrong and the opposite assumption will be correct. Taking this fact into account we can calculate the correctness as a value between 50 and 100% as follow:

$$\delta = 50\% + \left|50\% - \delta_i\right| \qquad (3)$$

**Fig. 4** shows the calculated correctness of the key candidates obtained by analyzing all 4 simulated traces.

As a reference we defined the worst-case of the attack result from the attacker's point of view as 50 per cent which means that the comparison to the mean method cannot even provide a slight hint whether the key bit processed is more likely a "1" or a "0", i.e. this means that the attack was not successful at all. The worst-case from the attacker's point of view is the ideal case from the designer's point of view.

Although we implemented our *kP* design strongly balanced and the field multiplier that is a kind of noise source is always active, the key was revealed with a correctness of 93 and 90 percent for the analysis of the synthesis and layout trace of the design *D_noUltra* respectively.

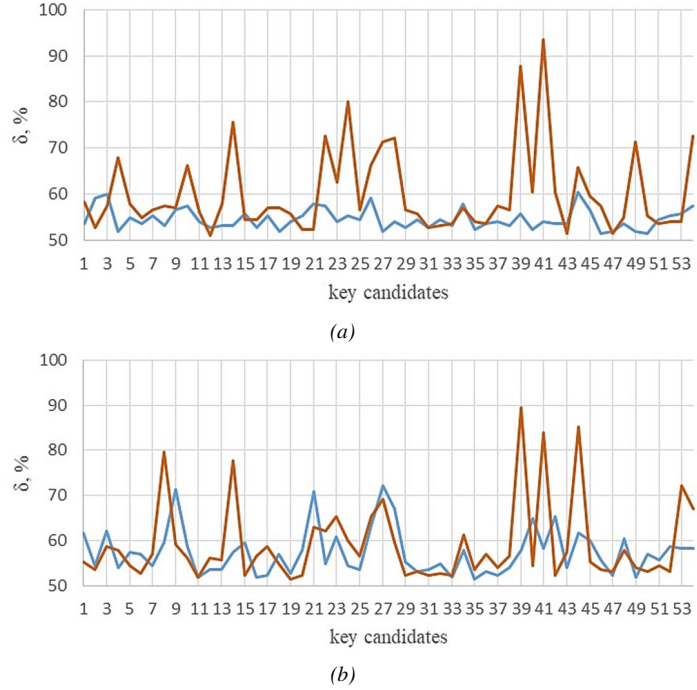

**Fig. 4.** Horizontal attack using *the the comparison to the mean* method: the analysis results are obtained using traces after synthesis *(a)* and layout *(b)* of the *kP* designs *D_noUltra* (orange) and *D_ultra* (blue) synthesized for the 250 nm technology.

The strong SCA leakage sources are the addressing of registers and the activity of the bus. Details are described in [21]. For the design *D_ultra*, i.e. for the same vhdl-code synthesized with option *compile_ultra* the best correctness for analysis of the power traces is 60 and 72 percent for synthesis and layout respectively. So, the optimization performed using the *compile_ultra* option hides information leakage significantly.

## 5    Produced ASIC

In order to understand how simulation results are comparable with results in real world, we produced the design *D_ultra* in the 250nm technology for a working frequency of 20 MHz (50ns clock cycle period as in the simulations) and bonded it to a printed circuit board (PCB). The PCB shown in **Fig. 5***(a)* contains 4 different *kP* designs for the NIST B-233 EC. We are focusing on ASIC01 which is the first die from the left (see **Fig. 5** *(b)*). The die has dimensions of 2458.36µm x 1086.12µm.

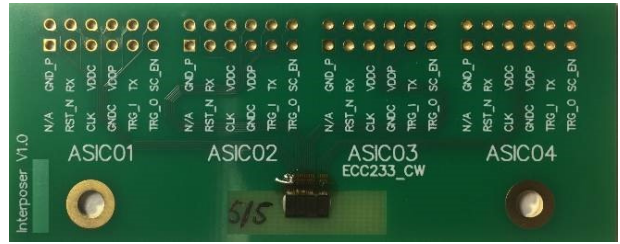

*(a)*

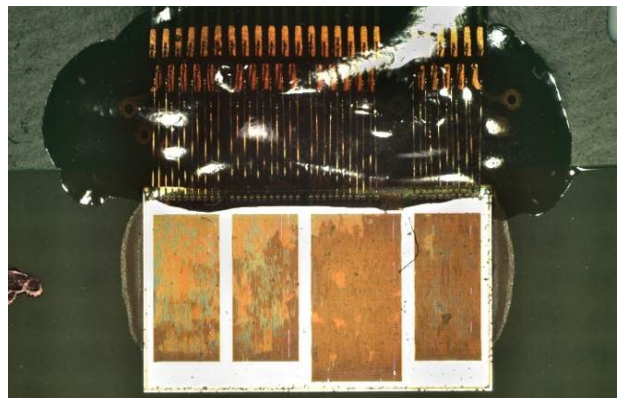

*(b)*

**Fig. 5.** A printed circuit board with produced ASIC *(a)* and a zoom of the unpackaged but bonded chips *(b).*

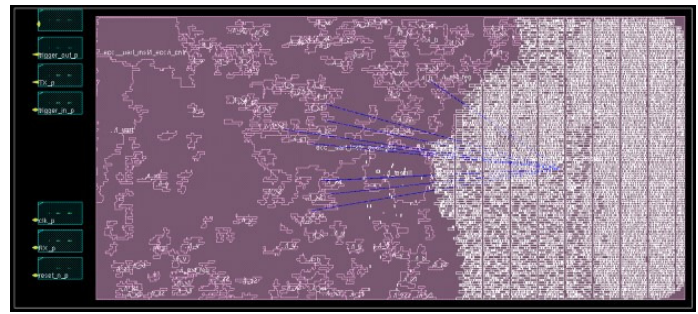

**Fig. 6.** Layout of ASIC01 obtained from Cadence Encounter. The white coloured area on the right hand side of the design coresponds to the multiplier block in the design.

The layout is shown on **Fig. 6**. The highlighted area – white coloured right hand side of the layout – shows shape and placement of the multiplier. The main parameters of the produced ASIC are given in **Table 2**.

**Table 2.** Parameters of the produced ASIC

| Parameter: | ASIC01 |
|---|---|
| Used Partial Multiplier | classical |
| Total area of Chip | 2670073.963 μm² |
| Chip Density (Total) | 91.345% |
| Chip Density (Subtracting Physical Cells) | 73.145% |
| Total area of Core | 2198931.840 μm² |
| Core Density (Total): | 99.592% |
| Total area of Standard cells | 2189956.608 μm² |
| Core Density (Subtracting Physical Cells) | 77.493% |
| Total area of Standard cells(Subtracting Physical Cells) | 1704016.94 μm² |
| Total area of Pad cells | 249011.112 μm² |

## 6    Horizontal DPA/DEMA attack against the ASIC

We captured power and electromagnetic traces of the *kP* execution on ASIC01 with a LeCroy WavePro 254HD oscilloscope (with a 2.5 GS/s sampling rate) using the Riscure current probe [32] and MFA-R 0.2-75 [33] near-field micro probe. The input data were the same as those we used to obtain simulation traces. Our measurement setup is shown in **Fig. 7**.

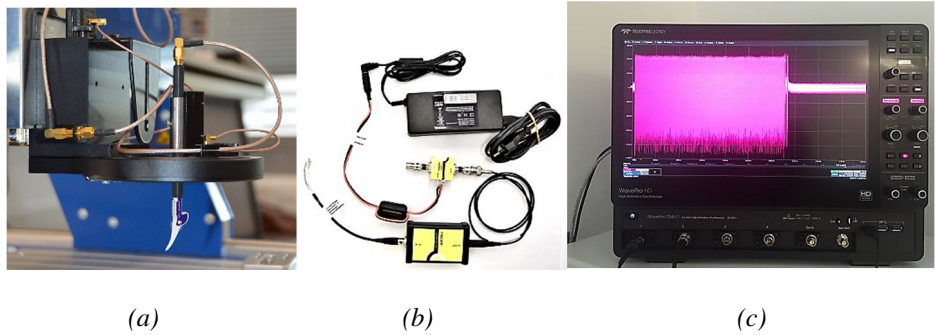

*(a)*                          *(b)*                          *(c)*

**Fig. 7.** Measurement setup: Riscure current probe *(a)*, Langer FLS-106 IC Scanner equipped with MFA-R 0.2-75 near-field micro probe *(b)* and LeCroy WavePro oscilloscope *(c)*.

We represented each clock cycle in the measured traces using only one value, that we calculated using all measured values (samples) within the clock cycle, i.e. we compressed the measured trace as follows:

$$v^{compressed} = \frac{1}{625} \cdot \sum_{l=1}^{625} \left(v_l^{measured}\right)^2 \qquad (4)$$

Here $v^{compressed}$ is the averaged squared amplitude value of all samples in a clock cycle; 625 is the number of measured values within the clock cycle. The squaring in (4) leads to the fact that the impact of the noise is significantly reduced. In addition the compression helps to reduce the complexity of the statistical analysis. The rest of attack is performed in the same way as described in section 3.

**Fig. 8** shows the results of the attacks against the design *D_ultra* design analysing its power traces (measured and simulated after layout) as well as attacking a measured electromagnetic trace.

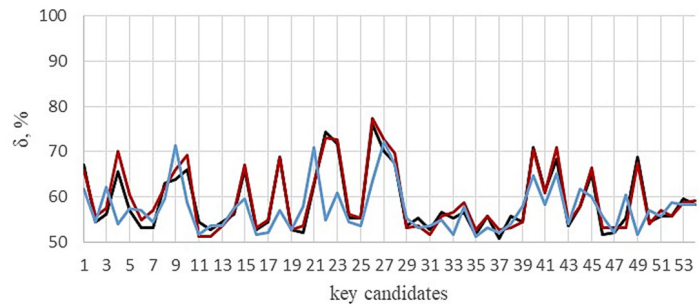

**Fig. 8.** Horizontal attacks using *the comparison to the mean* method: the *kP* design *D_ultra* produced in 250 nm technologie is attaked.

Black line in **Fig. 8** shows the attack results obtained analysing the measured power trace. The red line shows the attack results obtained analysing the measured electromagnetic trace. The blue line shows the attack results obtained analysing the simulated power trace obtained after layout.

## 7     Conclusions

This paper is to the best of our knowledge the first that detailed discussed the impact of different compile options on the success rate of side channel analysis attacks. In order to thoroughly investigate this impact we used two different compile options – *compile* and *compile_ultra*. We simulated power traces after synthesis and layout for both options. The *compile_ultra* option reduces the success rate significantly from 5 key candidates with a correctness between 75 and 90 per cent down to 3 key candidates with a maximum success rate of 72 per cent compared to the simple *compile* option. So we decided to manufacture the *kP* design with the *compile_ultra* version in our 250nm technology and then run the same attack against measured traces after getting the ASIC.

Although the success rate after layout shows a very high correlation with the one obtained running the attack against measured traces, there are some key candidates that are revealed with a higher correctness than in the simulation after layout. But still the success rate is by far better from a designer's point of view than what we expect for an ASIC manufactured with the *compile* option.

We also run the same attack against a measured electromagnetic trace and here we got a pretty high correlation of the success rate with the success rate even for simulated power traces. This is of high importance as detailed EM simulations for that large designs are currently infeasible.

Please note even though our results show that some compile options are helping to reduce side channel leakage while others don' t, compile options alone won' t make a bad design SCA resistant.